# Characterization of the Link Function in GN and EGN methods for Nonlinearity Assessment of Ultrawideband Coherent Fiber Optic Communication Systems with Raman Effect

**Mahdi Ranjbar Zefreh, Pierluigi Poggiolini**

*Politecnico di Torino, DET, Torino, Italy, mahdi.ranjbarzefreh@polito.it*

**Abstract:** In this paper, we present an accurate and numerically efficient method to implement the GN and EGN nonlinearity prediction methods when the power evolution along the fiber is in an arbitrary form. This approach will provide us with a reliable tool to efficiently use GN and Enhanced GN (EGN) methods for analysis of the nonlinearity for ultrawideband coherent fiber optic WDM systems (C+L band or even wider bandwidth systems) in the presence of the Inter-channel Stimulated Raman Scattering (ISRS) and systems using forward-pumped or/and backward-pumped Raman amplification.

## 1- Introduction

Physical-layer-aware control and optimization of ultra-high-capacity optical networks is becoming an increasingly important aspect of networking, as throughput demand and loads increase. A necessary pre-requisite is the availability of accurate analytical modeling of fiber non-linear effects (or NLI, Non-Linear-Interference).

Several NLI models have been proposed over the years, such as 'time-domain' [1], [2], GN [3], EGN [4] , [5] , others such as [6]-[9], and various precursors of the all of them (see for instance refs. in [10]). Among these methods, GN and EGN have been widely used to accurately evaluate the NLI in modern coherent optical communication systems.

On the other hand, increasing capacity demands is inevitable while deploying more new fibers is very costly and people are seeking methods to use the most available possible capacity of the already installed optical fibers. One important approach is extending the frequency bandwidth of the fiber from C band to C+L band and even higher. One of critical challenges in extending the bandwidth to C+L, specifically in the modeling point of view, is ISRS effect. ISRS is a nonlinear physical effect that causes the power transfer from higher frequencies of the propagating signal along the fiber to lower frequencies which is a minor neglectable effect in C band but becomes a dominant effect in systems using C+L bandwidth and higher bandwidths.

Also, coherent fiber optic system with Distributed Raman amplification are very favorable due to very low achievable noise figure and potential achievable high bandwidths [11] in Raman-amplified systems compared with EDFA based amplification systems. The nonlinearity modeling is an important issue in the Raman amplified fiber optic links.

The modeling of the NLI in the above-mentioned Raman-involved systems would be important for design and optimization of the system parameters. There are some good attempts to involve Raman effect in NLI modeling [12],[13],[14] however, they use several simplifying approximations which may cause inaccurate results in a general complex scenario, i.e. zero or low dispersion fiber involved fiber optic links or low Baud Rate involved channels in WDM comb. Therefore, the necessity of having an efficient tool for NLI modeling is obvious. As GN and EGN models have been very successful in accurate and effective modeling of the NLI for the Raman-free (C band) systems, they seem to be potentially attractive tools provided that, they can be equipped with Raman support capability.

The aim of this work is to reconsider the mathematical model of GN and EGN methods in the presence of the Raman effect. Then, we present a numerically efficient approach for the numerical implementation of the provided mathematical model. It is worth mentioning that in this paper we assume that we have the power evolution function of the signal in each WDM channel along the fiber and we use these power evolution functions in our model. The power evolution functions can be obtained by solving differential equations numerically which describe the Raman effect dynamism or other methods such as machine learning [15].

In section 2 we briefly explain the differential equations which model the power evolution of the channels when Raman effect is considered in the model. In section 3, a simple linear model of signal propagation along a fiber span is discussed. The *Link Function* (LF) in its general form is mathematically expressed in section 4. Sections 5 and 6 mainly discuss about efficient numerical implementation of the LF. In section 7, a step by step algorithm for practical implementation of the LF is presented and section 8 concludes the paper.

## 2- Raman effect mathematical modeling

Raman effect is a well-known nonlinear phenomenon which causes a power transfer from higher frequencies (shorter wavelengths) to lower frequencies (longer wavelengths) when signal propagates along a fiber optic. In a general scenario of a WDM system with possible Raman amplification, WDM frequency comb (signal) propagating in the +z direction in the fiber containing $N$ channels with center frequencies $f_1 < f_2 < \cdots < f_N$ and possible $N_p$ Raman amplification pumps (each pump can propagate either in +z direction or in -z direction) with frequencies $f_{N+1}, f_{N+2}, \ldots, f_{N+N_p}$, (all frequencies are positive absolute frequency) the distance dependent power of each channel and pump can be modeled by a set of coupled nonlinear differential equations as [16],[17]:

$$\varrho_l \times \frac{dP_l}{dz} = \left\{ \sum_{i=1}^{(N+N_p)} \zeta\left(\frac{f_l}{f_i}\right) \times C_R(f_i - f_l) \times P_i \right\} \times P_l - 2 \times \alpha_l \times P_l \qquad \text{eq. (1)}$$

$$; 1 \leq l \leq (N + N_p)$$

Where eq. (1) is a set of $(N + N_p)$ coupled nonlinear differential equations. $z$ is the distance from the signal at the signal launch location in the fiber and $+z$ is the signal propagation direction (forward propagating direction). $l$ can be in the range $1 \leq l \leq (N+N_p)$ and $P_j = P_j(z) = P(f_j, z)$ is the power of the $j$th WDM channel at the distance $z$ $(1 \leq j \leq N)$. $P_{j'+N} = P_{j'+N}(z) = P(f_{j'+N}, z)$ is the power of the $j'$th amplification pump at the distance $z$ $(1 \leq j' \leq N_p)$. $\varrho_l$ can be equal to either $+1$ or $-1$. $\varrho_l = +1$ for signal $(1 \leq l \leq N)$ or pumps propagating along $+z$ direction (forward propagating) while $\varrho_l = -1$ for pumps propagating in $-z$ direction (backward propagating pumps). Also, $\alpha_l$ is the intrinsic fiber loss parameter for the frequency $f_l$ in the absence of Raman which in general, can have different values for different frequencies (in general we have frequency dependent loss). We may also denote $\alpha_l$ by $\alpha(f_l)$ in this paper so $\alpha_l = \alpha(f_l)$. We assume the realistic assumption that $\alpha(f)$ is a slowly varying function with respect to the frequency and is approximately constant through each WDM channel while can be different from one channel to another one. $C_R(u)$ is an odd function with respect to its frequency variable $u$ which represents the gain profile of the Raman effect in the fiber which depends on the fiber physical specifications. For $u = 0, C_R(u) = 0$ and for $u > 0, C_R(u) \geq 0$ and for $u < 0, C_R(u) = -C_R(-u)$.

The function $\zeta(x)$ in equation (1) is defined as:

$$\zeta(x) \triangleq \begin{cases} x & x > 1 \\ 0 & x = 1 \\ 1 & x < 1 \end{cases} \qquad \text{eq. (2)}$$

It is worth noticing that in the absence of the Raman effect, $C_R(u) = 0\ \forall u$, and eq. (1) has the obvious analytical solution: $P_k(z) = P(f_k, z) = P(f_k, 0) \times e^{-2 \times \varrho_k \times \alpha_k \times z}$ for $1 \leq k \leq N + N_p$ where power of each channel is independent of the power of other channels.

In general, there is not an analytical solution for eq. (1), which then it must be solved numerically. In this work we assume that a pre-calculation has been made for obtaining power evolution of each WDM channel along each fiber span in the fiber optic link. Therefore, the function of power evolution for each WDM channel and for each fiber span has been already obtained and they are ready to be used in GN/EGN models.

### 3- Linear Model of Signal Propagation Along Fiber Span

A schematic figure of a fiber span with length $L_S^{(n_s)}$ is depicted in Fig. 1. We call the input signal (Electric field) to the fiber span as $E_{in}^{(n_s)}(f)$ in frequency domain. In the linear regime where we ignore Kerr nonlinearity and Raman effects and all other possible nonlinear effects in the fiber, the signal after propagating z meters along the fiber span is $E^{(n_s)}(z,f)$ in frequency domain and can be represented as:

$$E^{(n_s)}(z,f) = E_{in}^{(n_s)}(f) \times \exp\left(+\int_0^z \kappa^{(n_s)}(z',f)dz'\right) \qquad eq.\ (3)$$

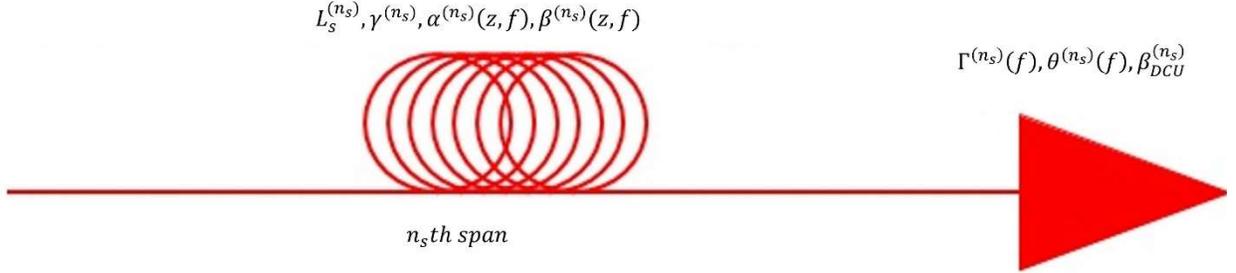

Figure (1): Mathematical model of a fiber span

Where $\kappa^{(n_s)}(z,f)$ is generalized propagation constant and is defined as:

$$\kappa^{(n_s)}(z,f) \triangleq -\alpha^{(n_s)}(z,f) - j\beta^{(n_s)}(z,f) = -\alpha^{(n_s)}(z,f) - j\beta^{(n_s)}(f) \qquad eq.\ (4)$$

Where $\alpha^{(n_s)}(z,f)$ and $\beta^{(n_s)}(z,f) = \beta^{(n_s)}(f)$ are two real valued functions namely: propagation loss function and propagation constant respectively. Combining equations (3), (4) the power evolution along the fiber can be written as:

$$P^{(n_s)}(z,f) \triangleq |E^{(n_s)}(z,f)|^2 = |E_{in}^{(n_s)}(f)|^2 \times \exp\left(-2 \times \int_0^z \alpha^{(n_s)}(z',f)dz'\right) = \qquad eq.\ (5)$$

$$P_{in}^{(n_s)}(f) \times \exp\left(-2 \times \int_0^z \alpha^{(n_s)}(z',f)dz'\right) = P^{(n_s)}(0,f) \times \exp\left(-2 \times \int_0^z \alpha^{(n_s)}(z',f)dz'\right)$$

For notation simplification and based on the equation (5) we define:

$$\rho^{(n_s)}(z,f) \triangleq \frac{P^{(n_s)}(z,f)}{P^{(n_s)}(0,f)} = \exp\left(-2 \times \int_0^z \alpha^{(n_s)}(z',f)dz'\right) \qquad eq.\ (6)$$

In our analysis we accept a realistic assumption which the power evolution for all fiber spans and for each frequency (each WDM channel) have been pre-calculated and therefore $\rho^{(n_s)}(z,f)$ for all WDM channel frequencies and for all spans in the optical link are available. It is worth mentioning that we realistically assume $\rho^{(n_s)}(z,f)$ is a slowly varying function with respect to $f$ and for each frequency inside one fixed channel in WDM comb it is approximately constant with respect to the frequency but can be changed channel by channel and can be different span by span. It is obvious that in the absence of Raman effect, $\rho^{(n_s)}(z,f) = \exp(-2 \times \alpha^{(n_s)}(f) \times z)$ where $\alpha^{(n_s)}(f)$ is the intrinsic fiber loss parameter of the $n_s$th fiber span in frequency $f$.

Another realistic approximation for modeling simplification would be $\beta^{(n_s)}(z,f) = \beta^{(n_s)}(f)$ which implies that the propagation constant is a constant function with respect to z (along a span) but can be a function of frequency and also can be different span by span. It is common to consider $\beta^{(n_s)}(f)$ with a Tylor expansion up to 4$^{th}$ term as:

$$\beta^{(n_s)}(z,f) = \beta^{(n_s)}(f)$$
$$= \beta_0^{(n_s)} + 2\pi \times \beta_1^{(n_s)} \times \left(f - f_c^{(n_s)}\right) + 4\pi^2 \times \frac{\beta_2^{(n_s)}}{2} \times \left(f - f_c^{(n_s)}\right)^2 \quad \text{eq. (7)}$$
$$+ 8\pi^3 \times \frac{\beta_3^{(n_s)}}{6} \times \left(f - f_c^{(n_s)}\right)^3$$

Where $f_c^{(n_s)}$ is the center frequency of the Tylor expansion of the propagation constant which is a constant parameter for each span while, in general, can be different span by span. Unlike $\alpha^{(n_s)}(z,f)$, $P^{(n_s)}(z,f)$ and $\rho^{(n_s)}(z,f)$ that are approximately constant functions along each channel, $\beta^{(n_s)}(f)$ is generally a rapidly varying function with respect to f and can vary along each channel.

Combining equations (4) and (7) we will have:

$$\kappa^{(n_s)}(z,f) \triangleq -\alpha^{(n_s)}(z,f) - j\beta_0^{(n_s)} - j2\pi \times \beta_1^{(n_s)} \times \left(f - f_c^{(n_s)}\right) \quad \text{eq. (8)}$$
$$-j4\pi^2 \times \frac{\beta_2^{(n_s)}}{2} \times \left(f - f_c^{(n_s)}\right)^2 - j\,8\pi^3 \times \frac{\beta_3^{(n_s)}}{6} \times \left(f - f_c^{(n_s)}\right)^3$$

The EDFA at the end of the span is also considered as a linear and time-invariant (LTI) system which ignoring the ASE noise and setting amplifier input signal in frequency domain as $E_{in,amp}^{(n_s)}(f)$, at the ouuput of amplifier we will have:

$$E_{out,amp}^{(n_s)}(f) = E_{in,amp}^{(n_s)}(f) \times \sqrt{\Gamma^{(n_s)}(f)} \times \quad \text{eq. (9)}$$
$$\exp\left(+j\theta^{(n_s)}(f)\right) \times \exp\left(-j \times 4\pi^2 \times \frac{\beta_{DCU}^{(n_s)}}{2} \times f^2\right)$$

In eq. (9), $\Gamma^{(n_s)}(f)$ is the power gain of the amplifier that generally can be frequency dependent. $\theta^{(n_s)}(f)$ is the frequency dependent phase which is imposed on the signal when passing through the amplifier and shows the possible linear filtering property of the EDFA amplifier. Also $\beta_{DCU}^{(n_s)}$ is the possible lumped accumulated dispersion integrated with EDFA amplifier. We also consider the realistic assumption for two functions $\Gamma^{(n_s)}(f)$ and $\theta^{(n_s)}(f)$ to be slowly varying functions with respect to the frequency and therefore, they are almost constant when f is changed through a fixed WDM channel but they can change channel by channel and also span by span.

As the GN and EGN are the perturbative models which consider the NLI as a small perturbation in the linear model of the signal, the linear model and parameters presented in this section show up in GN and EGN formulas extensively.

## 4- General Link Function

The GN and EGN formulas ae completely presented in [3], [4] for a coherent multi-span optical fiber link. All the specification of the optical fiber link contributes to GN and EGN formulas through the *Link Function* (LF). One general form of the LF is represented as [18]:

$$
\begin{aligned}
LK(f_1, f_2, f_3) = \\
-j \times \sum_{n_s=1}^{N_s} \gamma^{(n_s)} \times \left\{ \int_0^{L_S^{(n_s)}} e^{\int_0^{z'} \left[\kappa^{(n_s)}(z'', f_1) + \kappa^{(n_s)}(z'', f_2) + \left(\kappa^{(n_s)}(z'', f_3)\right)^* - \kappa^{(n_s)}(z'', f_1+f_2-f_3)\right] dz''} dz' \right\} \\
\times \prod_{p=n_s}^{N_s} \left\{ \sqrt{\Gamma^{(p)}(f_1+f_2-f_3)} \times e^{j\theta^{(p)}(f_1+f_2-f_3)} \times e^{\int_0^{L_S(p)} \kappa^{(p)}(z, f_1+f_2-f_3) dz} \, e^{-j\times 4\pi^2 \times \beta_{DCU}^{(p)} \times \frac{(f_1+f_2-f_3)^2}{2}} \right\} \\
\times \prod_{p=1}^{n_s-1} \left\{ \sqrt{\Gamma^{(p)}(f_1)\Gamma^{(p)}(f_2)\Gamma^{(p)}(f_3)} \times e^{\int_0^{L_S^{(p)}} \left[\kappa^{(p)}(z,f_1)+\kappa^{(p)}(z,f_2)+\left(\kappa^{(p)}(z,f_3)\right)^*\right] dz} \right. \\
\left. \times e^{j\left[\theta^{(p)}(f_1)+\theta^{(p)}(f_2)-\theta^{(p)}(f_3)\right]} \times e^{-j\times 4\pi^2 \times \left[\frac{\beta_{DCU}^{(p)}}{2} \times f_1^2 + \frac{\beta_{DCU}^{(p)}}{2} \times f_2^2 - \frac{\beta_{DCU}^{(p)}}{2} \times f_3^2\right]} \right\}
\end{aligned}
$$

eq. (10)

Where $N_s$ is the total spans that form the Optical fiber link and $L_S^{(n_s)}$ and $\gamma^{(n_s)}$ are the length and the nonlinearity parameter related to the $n_s$th fiber span respectively. Other parameters and variables used in equation (10) have been already presented in the previous section (section 3). The LF is denoted in chapter 7 of [19] by $\mu(f_1, f_2, f)$ notation which $f = f_1 + f_2 - f_3$. A less general (with respect to equation (10)) is presented in equation C.4 in [19]. GN formula was presented in equation (7.11) in [19] and EGN formulas are presented in equations (7.10-21) and (D.1-14).

Using equation (4), equation (10) can be simplified as:

$$
\begin{aligned}
LK(f_1, f_2, f_3) = -j \times e^{j\sum_{p=1}^{N_s} \theta^{(p)}(f_1+f_2-f_3)} \times e^{-j\times 4\pi^2 \times \frac{(f_1+f_2-f_3)^2}{2} \times \sum_{p=1}^{N_s} \beta_{DCU}^{(p)}} \\
\times e^{-j\times \sum_{p=1}^{N_s} L_S^{(p)} \times \beta^{(p)}(f_1+f_2-f_3)}
\end{aligned}
$$

eq. (11)

$$
\sum_{n_s=1}^{N_s} \gamma^{(n_s)} \times e^{j\sum_{p=1}^{(n_s-1)}\left[\theta^{(p)}(f_1)+\theta^{(p)}(f_2)-\theta^{(p)}(f_3)-\theta^{(p)}(f_1+f_2-f_3)\right]}
$$

$$
\times e^{-j\sum_{p=1}^{(n_s-1)} L_S^{(p)} \times \left[\beta^{(p)}(f_1)+\beta^{(p)}(f_2)-\beta^{(p)}(f_3)-\beta^{(p)}(f_1+f_2-f_3)\right]}
$$

$$
\times e^{-j\times 4\pi^2 \times \left[f_1^2+f_2^2-f_3^2-(f_1+f_2-f_3)^2\right] \times \sum_{p=1}^{(n_s-1)} \frac{\beta_{DCU}^{(p)}}{2}}
$$

$$
\times \left\{ \int_0^{L_S^{(n_s)}} e^{\int_0^{z'}\left[\kappa^{(n_s)}(z'', f_1)+\kappa^{(n_s)}(z'', f_2)+\left(\kappa^{(n_s)}(z'', f_3)\right)^*-\kappa^{(n_s)}(z'', f_1+f_2-f_3)\right] dz''} dz' \right\}
$$

$$
\times \prod_{p=n_s}^{N_s} \left\{ \sqrt{\Gamma^{(p)}(f_1+f_2-f_3)} \times e^{-\int_0^{L_S(p)} \alpha^{(p)}(z, f_1+f_2-f_3) dz} \right\} \times
$$

$$
\prod_{p=1}^{n_s-1} \left\{ \sqrt{\Gamma^{(p)}(f_1)\Gamma^{(p)}(f_2)\Gamma^{(p)}(f_3)} \times e^{-\int_0^{L_S(p)} \alpha^{(p)}(z, f_1) dz} e^{-\int_0^{L_S(p)} \alpha^{(p)}(z, f_2) dz} e^{-\int_0^{L_S(p)} \alpha^{(p)}(z, f_3) dz} \right\}
$$

Using equation (6), equation (11) can be written as:

$$LK(f_1, f_2, f_3) = -j \times e^{j \sum_{p=1}^{N_S} \theta^{(p)}(f_1+f_2-f_3)} \times e^{-j \times 4\pi^2 \times \frac{(f_1+f_2-f_3)^2}{2} \times \sum_{p=1}^{N_S} \beta_{DCU}^{(p)}} \quad eq.\ (12)$$

$$\times e^{-j \times \sum_{p=1}^{N_S} L_S^{(p)} \times \beta^{(p)}(f_1+f_2-f_3)}$$

$$\sum_{n_s=1}^{N_S} \gamma^{(n_s)} \times e^{j \sum_{p=1}^{(n_s-1)} [\theta^{(p)}(f_1)+\theta^{(p)}(f_2)-\theta^{(p)}(f_3)-\theta^{(p)}(f_1+f_2-f_3)]}$$

$$\times e^{-j \sum_{p=1}^{(n_s-1)} L_S^{(p)} \times [\beta^{(p)}(f_1)+\beta^{(p)}(f_2)-\beta^{(p)}(f_3)-\beta^{(p)}(f_1+f_2-f_3)]}$$

$$\times e^{-j \times 4\pi^2 \times [f_1^2 + f_2^2 - f_3^2 - (f_1+f_2-f_3)^2] \times \sum_{p=1}^{(n_s-1)} \frac{\beta_{DCU}^{(p)}}{2}}$$

$$\times \left\{ \int_0^{L_S^{(n_s)}} e^{\int_0^{z'} \left[ \kappa^{(n_s)}(z'', f_1) + \kappa^{(n_s)}(z'', f_2) + \left(\kappa^{(n_s)}(z'', f_3)\right)^* - \kappa^{(n_s)}(z'', f_1+f_2-f_3) \right] dz''} dz' \right\}$$

$$\times \prod_{p=n_s}^{N_S} \left\{ \sqrt{\Gamma^{(p)}(f_1+f_2-f_3) \times \rho^{(p)}\left(L_S^{(p)}, f_1+f_2-f_3\right)} \right\} \times$$

$$\prod_{p=1}^{n_s-1} \left\{ \sqrt{\Gamma^{(p)}(f_1)\Gamma^{(p)}(f_2)\Gamma^{(p)}(f_3) \times \rho^{(p)}\left(L_S^{(p)}, f_1\right) \times \rho^{(p)}\left(L_S^{(p)}, f_2\right) \times \rho^{(p)}\left(L_S^{(p)}, f_3\right)} \right\}$$

Using equation (7) and some manipulation in (12) we have:

$$LK(f_1, f_2, f_3) = -j \times e^{j \sum_{p=1}^{N_S} \theta^{(p)}(f_1+f_2-f_3)} \times e^{-j \times 4\pi^2 \times \frac{(f_1+f_2-f_3)^2}{2} \times \sum_{p=1}^{N_S} \beta_{DCU}^{(p)}} \quad eq.\ (13)$$

$$\times e^{-j \times \sum_{p=1}^{N_S} L_S^{(p)} \times \beta^{(p)}(f_1+f_2-f_3)}$$

$$\sum_{n_s=1}^{N_S} \gamma^{(n_s)} \times e^{j \sum_{p=1}^{(n_s-1)} [\theta^{(p)}(f_1)+\theta^{(p)}(f_2)-\theta^{(p)}(f_3)-\theta^{(p)}(f_1+f_2-f_3)]}$$

$$\times \prod_{p=n_s}^{N_S} \left\{ \sqrt{\Gamma^{(p)}(f_1+f_2-f_3) \times \rho^{(p)}\left(L_S^{(p)}, f_1+f_2-f_3\right)} \right\}$$

$$\times \prod_{p=1}^{(n_s-1)} \left\{ \sqrt{\Gamma^{(p)}(f_1)\Gamma^{(p)}(f_2)\Gamma^{(p)}(f_3) \times \rho^{(p)}\left(L_S^{(p)}, f_1\right) \times \rho^{(p)}\left(L_S^{(p)}, f_2\right) \times \rho^{(p)}\left(L_S^{(p)}, f_3\right)} \right\}$$

$$\times e^{+j \times 4\pi^2 \times (f_1-f_3) \times (f_2-f_3) \times \sum_{p=1}^{(n_s-1)} \left\{ \beta_{DCU}^{(p)} + L_S^{(p)} \times \left[ \beta_2^{(p)} + \pi \times \beta_3^{(p)} \times \left(f_1+f_2-2f_0^{(p)}\right) \right] \right\}}$$

$$\times \left\{ \int_0^{L_S^{(n_s)}} e^{\int_0^{z'} \left[ \kappa^{(n_s)}(z'', f_1) + \kappa^{(n_s)}(z'', f_2) + \left(\kappa^{(n_s)}(z'', f_3)\right)^* - \kappa^{(n_s)}(z'', f_1+f_2-f_3) \right] dz''} dz' \right\}$$

Also using equations (4), (6) and (7) we can see:

$$e^{\int_0^{z'}\left[\kappa^{(n_s)}(z'',\,f_1)+\kappa^{(n_s)}(z'',\,f_2)+\left(\kappa^{(n_s)}(z'',\,f_3)\right)^*-\kappa^{(n_s)}(z'',\,f_1+f_2-f_3)\right]dz''} \qquad eq.\ (14)$$

$$= e^{-\int_0^{z'}\alpha^{(n_s)}(z'',\,f_1)\,dz''} \times e^{-\int_0^{z'}\alpha^{(n_s)}(z'',\,f_2)\,dz''} \times e^{-\int_0^{z'}\alpha^{(n_s)}(z'',\,f_3)\,dz''}$$

$$\times e^{+\int_0^{z'}\alpha^{(n_s)}(z'',\,f_1+f_2-f_3)\,dz''}$$

$$\times e^{-j\times z'\times[\beta^{(n_s)}(f_1)+\beta^{(n_s)}(f_2)-\beta^{(n_s)}(f_3)-\beta^{(n_s)}(f_1+f_2-f_3)]}$$

$$= \sqrt{\frac{\rho^{(n_s)}(z',f_1)\times\rho^{(n_s)}(z',f_2)\times\rho^{(n_s)}(z',f_3)}{\rho^{(n_s)}(z',f_1+f_2-f_3)}}$$

$$\times e^{+j\times z'\times 4\pi^2\times(f_1-f_3)\times(f_2-f_3)\times[\beta_2^{(n_s)}+\pi\times\beta_3^{(n_s)}\times\left(f_1+f_2-2f_0^{(n_s)}\right)]}$$

Inserting equation (14) into equation (13) we have:

$$LK(f_1,f_2,f_3) = -j\times e^{j\sum_{p=1}^{N_S}\theta^{(p)}(f_1+f_2-f_3)}\times e^{-j\times 4\pi^2\times\frac{(f_1+f_2-f_3)^2}{2}\times\sum_{p=1}^{N_S}\beta_{DCU}^{(p)}} \qquad eq.\ (15)$$

$$\times e^{-j\times\sum_{p=1}^{N_S}L_S^{(p)}\times\beta^{(p)}(f_1+f_2-f_3)}\times$$

$$\sum_{n_s=1}^{N_S}\gamma^{(n_s)}\times e^{j\sum_{p=1}^{(n_s-1)}[\theta^{(p)}(f_1)+\theta^{(p)}(f_2)-\theta^{(p)}(f_3)-\theta^{(p)}(f_1+f_2-f_3)]}$$

$$\times \prod_{p=n_s}^{N_S}\left\{\sqrt{\Gamma^{(p)}(f_1+f_2-f_3)\times\rho^{(p)}\left(L_S^{(p)},f_1+f_2-f_3\right)}\right\}$$

$$\times \prod_{p=1}^{(n_s-1)}\left\{\sqrt{\Gamma^{(p)}(f_1)\Gamma^{(p)}(f_2)\Gamma^{(p)}(f_3)\times\rho^{(p)}\left(L_S^{(p)},f_1\right)\times\rho^{(p)}\left(L_S^{(p)},f_2\right)\times\rho^{(p)}\left(L_S^{(p)},f_3\right)}\right\}$$

$$\times e^{+j\times 4\pi^2\times(f_1-f_3)\times(f_2-f_3)\times\sum_{p=1}^{(n_s-1)}\left\{\beta_{DCU}^{(p)}+L_S^{(p)}\times\left[\beta_2^{(p)}+\pi\times\beta_3^{(p)}\times\left(f_1+f_2-2f_0^{(p)}\right)\right]\right\}}$$

$$\times \int_0^{L_S^{(n_s)}}\sqrt{\frac{\rho^{(n_s)}(z',f_1)\times\rho^{(n_s)}(z',f_2)\times\rho^{(n_s)}(z',f_3)}{\rho^{(n_s)}(z',f_1+f_2-f_3)}}$$

$$\times e^{+j\times z'\times 4\pi^2\times(f_1-f_3)\times(f_2-f_3)\times[\beta_2^{(n_s)}+\pi\times\beta_3^{(n_s)}\times\left(f_1+f_2-2f_0^{(n_s)}\right)]}\,dz'$$

In the absence of Raman effect where power evolution has a closed form solution as $\rho^{(n_s)}(z,f) = \exp(-2\times\alpha^{(n_s)}(f)\times z)$, equation (15) is simplified as:

$$LK(f_1,f_2,f_3) = -j \times e^{j\sum_{p=1}^{N_S} \theta^{(p)}(f_1+f_2-f_3)} \times e^{-j\times 4\pi^2 \times \frac{(f_1+f_2-f_3)^2}{2} \times \sum_{p=1}^{N_S} \beta_{DCU}^{(p)}}$$
$$\times e^{-j\times \sum_{p=1}^{N_S} L_S^{(p)} \times \beta^{(p)}(f_1+f_2-f_3)} \times$$

$$\sum_{n_s=1}^{N_S} \gamma^{(n_s)} \times e^{j\sum_{p=1}^{(n_s-1)}[\theta^{(p)}(f_1)+\theta^{(p)}(f_2)-\theta^{(p)}(f_3)-\theta^{(p)}(f_1+f_2-f_3)]}$$

$$\times \prod_{p=n_s}^{N_S} \left\{ \sqrt{\Gamma^{(p)}(f_1+f_2-f_3) \times \rho^{(p)}\left(L_S^{(p)}, f_1+f_2-f_3\right)} \right\}$$

$$\times \prod_{p=1}^{(n_s-1)} \left\{ \sqrt{\Gamma^{(p)}(f_1)\Gamma^{(p)}(f_2)\Gamma^{(p)}(f_3) \times \rho^{(p)}\left(L_S^{(p)}, f_1\right) \times \rho^{(p)}\left(L_S^{(p)}, f_2\right) \times \rho^{(p)}\left(L_S^{(p)}, f_3\right)} \right\}$$

$$\times e^{+j\times 4\pi^2 \times (f_1-f_3)\times(f_2-f_3)\times\sum_{p=1}^{(n_s-1)}\left\{\beta_{DCU}^{(p)}+L_S^{(p)}\times\left[\beta_2^{(p)}+\pi\times\beta_3^{(p)}\times\left(f_1+f_2-2f_0^{(p)}\right)\right]\right\}}$$

$$\times \int_0^{L_S^{(n_s)}} e^{-z' \times \left(\alpha^{(n_s)}(f_1)+\alpha^{(n_s)}(f_2)+\alpha^{(n_s)}(f_3)-\alpha^{(n_s)}(f_1+f_2-f_3)\right)}$$

$$\times e^{+j\times z' \times 4\pi^2 \times (f_1-f_3)\times(f_2-f_3)\times\left[\beta_2^{(n_s)}+\pi\times\beta_3^{(n_s)}\times\left(f_1+f_2-2f_0^{(n_s)}\right)\right]} dz'$$

eq. (16)

Where the last integral in the equation (16) has an analytic solution and therefore we will have:

$$LK(f_1,f_2,f_3) = -j \times e^{j\sum_{p=1}^{N_S} \theta^{(p)}(f_1+f_2-f_3)} \times e^{-j\times 4\pi^2 \times \frac{(f_1+f_2-f_3)^2}{2} \times \sum_{p=1}^{N_S} \beta_{DCU}^{(p)}}$$
$$\times e^{-j\times \sum_{p=1}^{N_S} L_S^{(p)} \times \beta^{(p)}(f_1+f_2-f_3)} \times$$

$$\sum_{n_s=1}^{N_S} \gamma^{(n_s)} \times e^{j\sum_{p=1}^{(n_s-1)}[\theta^{(p)}(f_1)+\theta^{(p)}(f_2)-\theta^{(p)}(f_3)-\theta^{(p)}(f_1+f_2-f_3)]}$$

$$\times \prod_{p=n_s}^{N_S} \left\{ \sqrt{\Gamma^{(p)}(f_1+f_2-f_3) \times \rho^{(p)}\left(L_S^{(p)}, f_1+f_2-f_3\right)} \right\}$$

$$\times \prod_{p=1}^{(n_s-1)} \left\{ \sqrt{\Gamma^{(p)}(f_1)\Gamma^{(p)}(f_2)\Gamma^{(p)}(f_3) \times \rho^{(p)}\left(L_S^{(p)}, f_1\right) \times \rho^{(p)}\left(L_S^{(p)}, f_2\right) \times \rho^{(p)}\left(L_S^{(p)}, f_3\right)} \right\}$$

$$\times e^{+j\times 4\pi^2 \times (f_1-f_3)\times(f_2-f_3)\times\sum_{p=1}^{(n_s-1)}\left\{\beta_{DCU}^{(p)}+L_S^{(p)}\times\left[\beta_2^{(p)}+\pi\times\beta_3^{(p)}\times\left(f_1+f_2-2f_0^{(p)}\right)\right]\right\}}$$

$$\times \frac{1 - e^{-L_S^{(n_s)} \times \left[\alpha^{(n_s)}(f_1)+\alpha^{(n_s)}(f_2)+\alpha^{(n_s)}(f_3)-\alpha^{(n_s)}(f_1+f_2-f_3) - j\times 4\pi^2 \times (f_1-f_3)\times(f_2-f_3)\times[\beta_2^{(n_s)}+\pi\times\beta_3^{(n_s)}\times(f_1+f_2-2f_0^{(n_s)})]\right]}}{\alpha^{(n_s)}(f_1)+\alpha^{(n_s)}(f_2)+\alpha^{(n_s)}(f_3)-\alpha^{(n_s)}(f_1+f_2-f_3) - j\times 4\pi^2 \times (f_1-f_3)\times(f_2-f_3)\times\left[\beta_2^{(n_s)}+\pi\times\beta_3^{(n_s)}\times\left(f_1+f_2-2f_0^{(n_s)}\right)\right]}$$

eq. (17)

But when Raman effect is present, $\rho^{(n_s)}(z,f)$ does not have an analytic solution and therefore the integral appeared in equation (15):

$$I^{(n_s)}(f_1, f_2, f_3) \triangleq \int_0^{L_s^{(n_s)}} \sqrt{\frac{\rho^{(n_s)}(z', f_1) \times \rho^{(n_s)}(z', f_2) \times \rho^{(n_s)}(z', f_3)}{\rho^{(n_s)}(z', f_1 + f_2 - f_3)}} \\ \times e^{+j \times z' \times 4\pi^2 \times (f_1 - f_3) \times (f_2 - f_3) \times [\beta_2^{(n_s)} + \pi \times \beta_3^{(n_s)} \times (f_1 + f_2 - 2f_0^{(n_s)})]} \, dz' \qquad eq.\ (18)$$

must be numerically calculated. For notation simplicity we define:

$$\vartheta^{(n_s)}(z', f_1, f_2, f_3) \triangleq \sqrt{\frac{\rho^{(n_s)}(z', f_1) \times \rho^{(n_s)}(z', f_2) \times \rho^{(n_s)}(z', f_3)}{\rho^{(n_s)}(z', f_1 + f_2 - f_3)}} \qquad eq.\ (19)$$

$$\delta^{(n_s)}(f_1, f_2, f_3) \\ \triangleq j \times 4\pi^2 \times (f_1 - f_3) \times (f_2 - f_3) \times \left[\beta_2^{(n_s)} + \pi \times \beta_3^{(n_s)} \times \left(f_1 + f_2 - 2f_0^{(n_s)}\right)\right] \qquad eq.\ (20)$$

Using two above definitions in equations (19) and (20), equation (18) is expressed as:

$$I^{(n_s)}(f_1, f_2, f_3) \triangleq \int_0^{L_s^{(n_s)}} \vartheta^{(n_s)}(z', f_1, f_2, f_3) \times e^{+z' \times \delta^{(n_s)}(f_1, f_2, f_3)} \, dz' \qquad eq.\ (21)$$

We also define:

$$\begin{aligned}\Psi^{(n_s)}(z', f_1, f_2, f_3) &\triangleq \vartheta^{(n_s)}(z', f_1, f_2, f_3) \times \\ &\quad exp\bigl(z' \times [\alpha^{(n_s)}(f_1) + \alpha^{(n_s)}(f_2) + \alpha^{(n_s)}(f_3) - \alpha^{(n_s)}(f_1 + f_2 - f_3)]\bigr) \\ &= \sqrt{\frac{\rho^{(n_s)}(z', f_1) \times \rho^{(n_s)}(z', f_2) \times \rho^{(n_s)}(z', f_3)}{\rho^{(n_s)}(z', f_1 + f_2 - f_3)}} \\ &\quad \times exp\bigl(z' \times [\alpha^{(n_s)}(f_1) + \alpha^{(n_s)}(f_2) + \alpha^{(n_s)}(f_3) - \alpha^{(n_s)}(f_1 + f_2 - f_3)]\bigr)\end{aligned} \qquad eq.\ (22)$$

$$\begin{aligned}\Theta^{(n_s)}(f_1, f_2, f_3) &\triangleq -\alpha^{(n_s)}(f_1) - \alpha^{(n_s)}(f_2) - \alpha^{(n_s)}(f_3) + \alpha^{(n_s)}(f_1 + f_2 - f_3) \\ &\quad + j4\pi^2 \times (f_1 - f_3) \times (f_2 - f_3) \times \left[\beta_2^{(n_s)} + \pi \times \beta_3^{(n_s)} \times \left(f_1 + f_2 - 2f_0^{(n_s)}\right)\right]\end{aligned} \qquad eq.\ (23)$$

Where $\Theta^{(n_s)}(f_1, f_2, f_3)$ is a z-independent function. Using (22) and (23), the integral in equation (21) can be written as:

$$I^{(n_s)}(f_1, f_2, f_3) = \int_0^{L_s^{(n_s)}} \Psi^{(n_s)}(z', f_1, f_2, f_3) \times e^{+z' \Theta^{(n_s)}(f_1, f_2, f_3)} \, dz' \qquad eq.\ (24)$$

Using equation (18), LF in equation (15) can be rewritten as:

$$LK(f_1,f_2,f_3) = -j \times e^{j\sum_{p=1}^{N_S} \theta^{(p)}(f_1+f_2-f_3)} \times e^{-j\times 4\pi^2 \times \frac{(f_1+f_2-f_3)^2}{2} \times \sum_{p=1}^{N_S} \beta_{DCU}^{(p)}} \qquad eq.\ (25)$$
$$\times e^{-j\times\sum_{p=1}^{N_S} L_S^{(p)} \times \beta^{(p)}(f_1+f_2-f_3)} \times$$
$$\sum_{n_s=1}^{N_S} \gamma^{(n_s)} \times e^{j\sum_{p=1}^{(n_s-1)}[\theta^{(p)}(f_1)+\theta^{(p)}(f_2)-\theta^{(p)}(f_3)-\theta^{(p)}(f_1+f_2-f_3)]}$$
$$\times \prod_{p=n_s}^{N_S} \left\{ \sqrt{\Gamma^{(p)}(f_1+f_2-f_3) \times \rho^{(p)}\left(L_S^{(p)}, f_1+f_2-f_3\right)} \right\}$$
$$\times \prod_{p=1}^{(n_s-1)} \left\{ \sqrt{\Gamma^{(p)}(f_1)\Gamma^{(p)}(f_2)\Gamma^{(p)}(f_3)} \times \rho^{(p)}\left(L_S^{(p)}, f_1\right) \times \rho^{(p)}\left(L_S^{(p)}, f_2\right) \times \rho^{(p)}\left(L_S^{(p)}, f_3\right) \right\}$$
$$\times e^{+j\times 4\pi^2 \times (f_1-f_3)\times(f_2-f_3)\times \sum_{p=1}^{(n_s-1)}\left\{\beta_{DCU}^{(p)}+L_S^{(p)}\times\left[\beta_2^{(p)}+\pi\times\beta_3^{(p)}\times\left(f_1+f_2-2f_0^{(p)}\right)\right]\right\}} \times I^{(n_s)}(f_1,f_2,f_3)$$

## 5- Numerical Implementation of Integral $I^{(n_s)}(f_1,f_2,f_3)$

The integral $I^{(n_s)}(f_1,f_2,f_3)$ has the analytic solution when the Raman effect is ignored (presented in equation (17)) but in the general scenario where Raman effect is present, $I^{(n_s)}(f_1,f_2,f_3)$ must be numerically evaluated.

For numerically implementation of $I^{(n_s)}(f_1,f_2,f_3)$ directly, we have several problems. It is worth mentioning that in the GN model we have a double integral on two dimensional $f_1-f_2$ plane while adding another integral on z variable changes the numerical integration to a more complex three dimensional in $f_1-f_2-z$ space. Furthermore, to implement the integral in equation (18) numerically with acceptable accuracy, due to the constant amplitude but phase varying term $e^{+z'\times\delta^{(n_s)}(f_1,f_2,f_3)}$, the needed resolution of integral in $z'$ is very dependent to the absolute value of the term $\delta^{(n_s)}(f_1,f_2,f_3)$. In fact, $\delta^{(n_s)}(f_1,f_2,f_3)$ which is dependent to values of $f_1,f_2,f_3$ can change in a wide range specifically for wide band systems and as a result the z-step integration adjusting would be tricky. Therefore, the goal of this section is to present an approach to avoid direct integration for evaluation of $I^{(n_s)}(f_1,f_2,f_3)$.

Suppose, if $\Psi^{(n_s)}(z',f_1,f_2,f_3)$ in equation (24) has a finite degree polynomial representation (with respect to $z'$ variable), $I^{(n_s)}(f_1,f_2,f_3)$ will have an analytic representation. For now, we assume the hypothetical finite polynomial representation of $\Psi^{(n_s)}(z',f_1,f_2,f_3)$ as:

$$\Psi^{(n_s)}(z',f_1,f_2,f_3) = \sum_{k=0}^{N_\Psi} h_k^{(n_s)}(f_1,f_2,f_3) \times z'^k \qquad eq.\ (26)$$

Where $N_\Psi$ is the polynomial degree and $h_k^{(n_s)}(f_1,f_2,f_3)$ is the $z'$-independent coefficient of the $z'^k$ term in the polynomial representation of $\Psi^{(n_s)}(z',f_1,f_2,f_3)$. Using (26), the integral in equation (24) becomes:

$$I^{(n_s)}(f_1, f_2, f_3) = \sum_{k=0}^{N_\Psi} h_k^{(n_s)}(f_1, f_2, f_3) \times \int_0^{L_s^{(n_s)}} z'^k \times e^{+z'\Theta^{(n_s)}(f_1,f_2,f_3)} \, dz' \qquad eq.\ (27)$$

Since $\Theta^{(n_s)}(f_1, f_2, f_3)$ is a z-independent function, the integral in equation (27) has an analytic solution:

$$\int z'^k \times e^{+z'\Theta^{(n_s)}(f_1,f_2,f_3)} \, dz' = \frac{e^{+z \times \Theta^{(n_s)}(f_1,f_2,f_3)}}{[\Theta^{(n_s)}(f_1,f_2,f_3)]^{(k+1)}} \times \sum_{m=0}^{k} \frac{(-1)^m \times k! \times [\Theta^{(n_s)}(f_1,f_2,f_3)]^{(k-m)}}{(k-m)!} z^{(k-m)} \qquad eq.\ (28)$$

Therefore, using (28), the result of integral in equation (27) will be:

$$\int_0^{L_s^{(n_s)}} z'^k \times e^{+z'\Theta^{(n_s)}(f_1,f_2,f_3)} \, dz' = \frac{e^{+L_s^{(n_s)} \times \Theta^{(n_s)}(f_1,f_2,f_3)}}{[\Theta^{(n_s)}(f_1,f_2,f_3)]^{(k+1)}} \times \sum_{m=0}^{k} \frac{(-1)^m \times k! \times [\Theta^{(n_s)}(f_1,f_2,f_3)]^{(k-m)}}{(k-m)!} \times \left[L_s^{(n_s)}\right]^{(k-m)} - \frac{(-1)^k \times k!}{[\Theta^{(n_s)}(f_1,f_2,f_3)]^{(k+1)}} \qquad eq.\ (29)$$

Using equation (29), The integral in equation (27) is represented as:

$$I^{(n_s)}(f_1, f_2, f_3) = \sum_{k=0}^{N_\Psi} h_k^{(n_s)}(f_1, f_2, f_3) \times \frac{e^{+L_s^{(n_s)} \times \Theta^{(n_s)}(f_1,f_2,f_3)}}{[\Theta^{(n_s)}(f_1,f_2,f_3)]^{(k+1)}}$$
$$\times \sum_{m=0}^{k} \frac{(-1)^m \times k! \times [\Theta^{(n_s)}(f_1,f_2,f_3)]^{(k-m)}}{(k-m)!} \times \left[L_s^{(n_s)}\right]^{(k-m)}$$
$$- \sum_{k=0}^{N_\Psi} h_k^{(n_s)}(f_1, f_2, f_3) \times \frac{(-1)^k \times k!}{[\Theta^{(n_s)}(f_1,f_2,f_3)]^{(k+1)}} \qquad eq.\ (30)$$

Manipulating equation (30) we will have:

$$I^{(n_s)}(f_1,f_2,f_3) = \sum_{m=0}^{N_\Psi} \frac{e^{+L_S^{(n_s)} \times \Theta^{(n_s)}(f_1,f_2,f_3)}}{[\Theta^{(n_s)}(f_1,f_2,f_3)]^{(m+1)}} \times \sum_{k=m}^{N_\Psi} \frac{(-1)^m \times h_k^{(n_s)}(f_1,f_2,f_3) \times k! \times}{(k-m)!}$$
$$\times \left[L_S^{(n_s)}\right]^{(k-m)}$$
$$- \sum_{m=0}^{N_\Psi} h_m^{(n_s)}(f_1,f_2,f_3) \times \frac{(-1)^m \times m!}{[\Theta^{(n_s)}(f_1,f_2,f_3)]^{(m+1)}}$$

eq. (31)

For notation simplicity we define:

$$\tau_m^{(n_s)}(f_1,f_2,f_3) \triangleq \sum_{k=m}^{N_\Psi} \frac{(-1)^m \times h_k^{(n_s)}(f_1,f_2,f_3) \times k! \times}{(k-m)!} \times \left[L_S^{(n_s)}\right]^{(k-m)}$$

eq. (32)

Therefore, using equation (32), equation (31) can be written as:

$$I^{(n_s)}(f_1,f_2,f_3)$$
$$= \sum_{m=0}^{N_\Psi} \frac{e^{+L_S^{(n_s)} \times \Theta^{(n_s)}(f_1,f_2,f_3)} \times \tau_m^{(n_s)}(f_1,f_2,f_3) - (-1)^m \times m! \times h_m^{(n_s)}(f_1,f_2,f_3)}{[\Theta^{(n_s)}(f_1,f_2,f_3)]^{(m+1)}}$$

eq. (33)

Using (33), the LF in equation (25) is written as:

$$LK(f_1,f_2,f_3) = -j \times e^{j\sum_{p=1}^{N_S}\theta^{(p)}(f_1+f_2-f_3)} \times e^{-j\times 4\pi^2 \times \frac{(f_1+f_2-f_3)^2}{2} \times \sum_{p=1}^{N_S}\beta_{DCU}^{(p)}}$$
$$\times e^{-j\times\sum_{p=1}^{N_S}L_S^{(p)}\times\beta^{(p)}(f_1+f_2-f_3)} \times$$
$$\sum_{n_s=1}^{N_S} \gamma^{(n_s)} \times e^{j\sum_{p=1}^{(n_s-1)}[\theta^{(p)}(f_1)+\theta^{(p)}(f_2)-\theta^{(p)}(f_3)-\theta^{(p)}(f_1+f_2-f_3)]}$$
$$\times \prod_{p=n_s}^{N_S}\left\{\sqrt{\Gamma^{(p)}(f_1+f_2-f_3) \times \rho^{(p)}\left(L_S^{(p)},f_1+f_2-f_3\right)}\right\}$$
$$\times \prod_{p=1}^{(n_s-1)}\left\{\sqrt{\Gamma^{(p)}(f_1)\Gamma^{(p)}(f_2)\Gamma^{(p)}(f_3) \times \rho^{(p)}\left(L_S^{(p)},f_1\right) \times \rho^{(p)}\left(L_S^{(p)},f_2\right) \times \rho^{(p)}\left(L_S^{(p)},f_3\right)}\right\}$$
$$\times e^{+j\times 4\pi^2 \times (f_1-f_3)\times(f_2-f_3)\times\sum_{p=1}^{(n_s-1)}\left\{\beta_{DCU}^{(p)}+L_S^{(p)}\times\left[\beta_2^{(p)}+\pi\times\beta_3^{(p)}\times\left(f_1+f_2-2f_0^{(p)}\right)\right]\right\}}$$
$$\times \sum_{m=0}^{N_\Psi} \frac{e^{+L_S^{(n_s)}\times\Theta^{(n_s)}(f_1,f_2,f_3)} \times \tau_m^{(n_s)}(f_1,f_2,f_3) - (-1)^m \times m! \times h_m^{(n_s)}(f_1,f_2,f_3)}{[\Theta^{(n_s)}(f_1,f_2,f_3)]^{(m+1)}}$$

eq. (34)

All the terms behind the sign $\sum_{n_s=1}^{N_s}$ in equation (34) can be written as $e^{j\varphi(f_1+f_2-f_3)}$ where the $\varphi(.)$ is a real value function. It can be mathematically shown that the aforementioned function $(e^{j\varphi(f_1+f_2-f_3)})$ is ineffective in the final result of GN and EGN formulas and can be eliminated (replaced by 1) without any change in GN or EGN results. It is due to fact that $f = f_1 + f_2 - f_3$ and the mathematical form of GN and EGN formulas presented in equations (7.10-21) and (D.1-14) in [19]. Therefore, LF can be more simply represented as:

$$LK(f_1, f_2, f_3) =$$
$$\sum_{n_s=1}^{N_s} \gamma^{(n_s)} \times e^{j\sum_{p=1}^{(n_s-1)}[\theta^{(p)}(f_1)+\theta^{(p)}(f_2)-\theta^{(p)}(f_3)-\theta^{(p)}(f_1+f_2-f_3)]}$$
$$\times \prod_{p=n_s}^{N_s} \left\{ \sqrt{\Gamma^{(p)}(f_1+f_2-f_3)} \times \rho^{(p)}\left(L_s^{(p)}, f_1+f_2-f_3\right) \right\}$$
$$\times \prod_{p=1}^{(n_s-1)} \left\{ \sqrt{\Gamma^{(p)}(f_1)\Gamma^{(p)}(f_2)\Gamma^{(p)}(f_3)} \times \rho^{(p)}\left(L_s^{(p)}, f_1\right) \times \rho^{(p)}\left(L_s^{(p)}, f_2\right) \times \rho^{(p)}\left(L_s^{(p)}, f_3\right) \right\}$$
$$\times e^{+j\times 4\pi^2 \times (f_1-f_3)\times(f_2-f_3)\times\sum_{p=1}^{(n_s-1)}\{\beta_{DCU}^{(p)}+L_s^{(p)}\times[\beta_2^{(p)}+\pi\times\beta_3^{(p)}\times(f_1+f_2-2f_0^{(p)})]\}}$$
$$\times \sum_{m=0}^{N_\Psi} \frac{e^{+L_s^{(n_s)}\times\theta^{(n_s)}(f_1,f_2,f_3)} \times \tau_m^{(n_s)}(f_1,f_2,f_3) - (-1)^m \times m! \times h_m^{(n_s)}(f_1,f_2,f_3)}{[\theta^{(n_s)}(f_1,f_2,f_3)]^{(m+1)}}$$

eq. (34.2)

## 6- Optimum Calculation of the Coefficients $h_m^{(n_s)}(f_1,f_2,f_3)$

As we mentioned in the previous sections, the integral $I^{(n_s)}(f_1, f_2, f_3)$ in the LF does not have an analytic solution in general however, as we saw in the previous section, we can obtain an analytic solution provided that, $\Psi^{(n_s)}(z', f_1, f_2, f_3)$ can be represented as a finite degree polynomial as in equation (26). In general, we only have the numerical values of $\Psi^{(n_s)}(z', f_1, f_2, f_3)$ by solving set of differential equations presented in equation (1) but we try to fit them to a polynomial. Therefore, we try to find $h_m^{(n_s)}(f_1, f_2, f_3)$ values to make the below cost function minimum:

$$C^{(n_s)}(f_1,f_2,f_3) \triangleq \int_0^{L_s^{(n_s)}} W_0(z') \times \left[\frac{\Psi^{(n_s)}(z',f_1,f_2,f_3) - \sum_{k=0}^{N_\Psi} h_k^{(n_s)}(f_1,f_2,f_3) \times z'^k}{\Psi^{(n_s)}(z',f_1,f_2,f_3)}\right]^2 dz' \quad \text{eq. (35)}$$

Where $W_0(z')$ in equation (35) is a weigh function. We consider the weight function as $W_0(z') = [\vartheta^{(n_s)}(z', f_1, f_2, f_3)]^{m_W}$ where $m_W$ is a positive constant which gives one degree of freedom in our analysis. The reason behind selection of this weight function is that our goal is to calculate the integral presented in equation (21) accurately. As $\times \delta^{(n_s)}(f_1, f_2, f_3)$ is a pure imaginary function, we have $\left|e^{+z'\times\delta^{(n_s)}(f_1,f_2,f_3)}\right| = 1; \forall z'$. Therefore, the greater values of the function $\vartheta^{(n_s)}(z', f_1, f_2, f_3)$, which is a real and positive function, have more contribution in the integral. In the other words, the major part of nonlinearity is made at higher powers and therefore we need to be more accurate in higher powers which have more effect on NLI generation.

Therefore, the cost function in equation (35) is given by:

$$C^{(n_s)}(f_1, f_2, f_3) \triangleq \int_0^{L_s^{(n_s)}} W(z') \times \left[ \Psi^{(n_s)}(z', f_1, f_2, f_3) - \sum_{k=0}^{N_\Psi} h_k^{(n_s)}(f_1, f_2, f_3) \times z'^k \right]^2 dz' \quad \text{eq. (36)}$$

Where $W(z')$ in equation (36) is:

$$W(z') \triangleq \frac{\left[ \vartheta^{(n_s)}(z', f_1, f_2, f_3) \right]^{m_W}}{\left[ \Psi^{(n_s)}(z', f_1, f_2, f_3) \right]^2} \quad \text{eq. (37)}$$

To minimize the cost function $C^{(n_s)}(f_1, f_2, f_3)$ and find the optimum values of $h_k^{(n_s)}(f_1, f_2, f_3)$ we should have:

$$\frac{\partial C^{(n_s)}(f_1, f_2, f_3)}{\partial h_k^{(n_s)}(f_1, f_2, f_3)} = 0 \quad ; \forall k = 0, 1, 2, \ldots, N_\Psi \quad \text{eq. (38)}$$

Taking derivative from both sides of equation (36):

$$\frac{\partial C^{(n_s)}(f_1, f_2, f_3)}{\partial h_k^{(n_s)}(f_1, f_2, f_3)} = \int_0^{L_s^{(n_s)}} W(z') \times 2 \times [-z'^n] \times \left[ \Psi^{(n_s)}(z', f_1, f_2, f_3) - \sum_{k=0}^{N_\Psi} h_k^{(n_s)}(f_1, f_2, f_3) \times z'^k \right] dz' = 0 \quad ; \forall n = 0, 1, 2, \ldots, N_\Psi \quad \text{eq. (39)}$$

(39) leads to a set of $(N_\Psi + 1)$ of unknowns with $(N_\Psi + 1)$ equations as the matrix equation below:

$$\begin{bmatrix} \int_0^{L_s^{(n_s)}} W(z') \times \Psi^{(n_s)}(z', f_1, f_2, f_3) dz' & \int_0^{L_s^{(n_s)}} W(z') \times z' \times \Psi^{(n_s)}(z', f_1, f_2, f_3) dz' & \ldots & \int_0^{L_s^{(n_s)}} W(z') \times z'^{N_\Psi} \times \Psi^{(n_s)}(z', f_1, f_2, f_3) dz' \\ \int_0^{L_s^{(n_s)}} W(z') \times z' \times \Psi^{(n_s)}(z', f_1, f_2, f_3) dz' & \int_0^{L_s^{(n_s)}} W(z') \times z'^2 \times \Psi^{(n_s)}(z', f_1, f_2, f_3) dz' & \ldots & \int_0^{L_s^{(n_s)}} W(z') \times z'^{(N_\Psi+1)} \times \Psi^{(n_s)}(z', f_1, f_2, f_3) dz' \\ \vdots & \vdots & \vdots & \vdots \\ \int_0^{L_s^{(n_s)}} W(z') \times z'^{N_\Psi} \times \Psi^{(n_s)}(z', f_1, f_2, f_3) dz' & \int_0^{L_s^{(n_s)}} W(z') \times z'^{(N_\Psi+1)} \times \Psi^{(n_s)}(z', f_1, f_2, f_3) dz' & \ldots & \int_0^{L_s^{(n_s)}} W(z') \times z'^{(2N_\Psi)} \times \Psi^{(n_s)}(z', f_1, f_2, f_3) dz' \end{bmatrix}$$

$$\times \begin{bmatrix} h_0^{(n_s)}(f_1, f_2, f_3) \\ h_1^{(n_s)}(f_1, f_2, f_3) \\ \vdots \\ h_{N_\Psi}^{(n_s)}(f_1, f_2, f_3) \end{bmatrix} = \begin{bmatrix} \int_0^{L_s^{(n_s)}} W(z') \times \Psi^{(n_s)}(z', f_1, f_2, f_3) dz' \\ \int_0^{L_s^{(n_s)}} W(z') \times z' \times \Psi^{(n_s)}(z', f_1, f_2, f_3) dz' \\ \vdots \\ \int_0^{L_s^{(n_s)}} W(z') \times z'^{N_\Psi} \times \Psi^{(n_s)}(z', f_1, f_2, f_3) dz' \end{bmatrix} \quad \text{eq. (40)}$$

Therefore, we have:

$$\begin{bmatrix} h_0^{(n_s)}(f_1,f_2,f_3) \\ h_1^{(n_s)}(f_1,f_2,f_3) \\ \vdots \\ \vdots \\ h_{N_\Psi}^{(n_s)}(f_1,f_2,f_3) \end{bmatrix} = \begin{bmatrix} \int_0^{L_s^{(n_s)}} W(z')\times\Psi^{(n_s)}(z',f_1,f_2,f_3)dz' & \int_0^{L_s^{(n_s)}} W(z')\times z'\times\Psi^{(n_s)}(z',f_1,f_2,f_3)dz' & \cdots\cdots & \int_0^{L_s^{(n_s)}} W(z')\times z'^{N_\Psi}\times\Psi^{(n_s)}(z',f_1,f_2,f_3)dz' \\ \int_0^{L_s^{(n_s)}} W(z')\times z'\times\Psi^{(n_s)}(z',f_1,f_2,f_3)dz' & \int_0^{L_s^{(n_s)}} W(z')\times z'^{2}\times\Psi^{(n_s)}(z',f_1,f_2,f_3)dz' & \cdots\cdots & \int_0^{L_s^{(n_s)}} W(z')\times z'^{(N_\Psi+1)}\times\Psi^{(n_s)}(z',f_1,f_2,f_3)dz' \\ \vdots & \vdots & \vdots & \vdots \\ \int_0^{L_s^{(n_s)}} W(z')\times z'^{N_\Psi}\times\Psi^{(n_s)}(z',f_1,f_2,f_3)dz' & \int_0^{L_s^{(n_s)}} W(z')\times z'^{(N_\Psi+1)}\times\Psi^{(n_s)}(z',f_1,f_2,f_3)dz' & \cdots\cdots & \int_0^{L_s^{(n_s)}} W(z')\times z'^{(2N_\Psi)}\times\Psi^{(n_s)}(z',f_1,f_2,f_3)dz' \end{bmatrix}^{-1}$$

$$\times \begin{bmatrix} \int_0^{L_s^{(n_s)}} W(z')\times\Psi^{(n_s)}(z',f_1,f_2,f_3)dz' \\ \int_0^{L_s^{(n_s)}} W(z')\times z'\times\Psi^{(n_s)}(z',f_1,f_2,f_3)dz' \\ \vdots \\ \int_0^{L_s^{(n_s)}} W(z')\times z'^{N_\Psi}\times\Psi^{(n_s)}(z',f_1,f_2,f_3)dz' \end{bmatrix}$$

*eq. (41)*

The values of $h_k^{(n_s)}(f_1,f_2,f_3)$ are calculated by equation (41) once for each span and for each frequency integration island which is defined in [20]. But inside an integration island the values of $h_k^{(n_s)}(f_1,f_2,f_3)$ remain constant and this is exactly the reason we followed the above explained approach. Therefore, inside a frequency integration island all numerical integrals inside the matrix equation (41) can be calculated by with a constant $\Delta z$ (integration step) which $\Delta z$ does not depend on the range of $f_1, f_2, f_3$ due to integration island. On the contrary if we wanted to accurately, directly and numerically calculate the $I^{(n_s)}(f_1,f_2,f_3)$ in equation (18), $\Delta z$ (integration step) would be strongly depend to the range of $f_1, f_2, f_3$ in the frequency integration island and it made us to calculate complicated triple integrals over three-dimensional spaces $(f_1 - f_2 - z)$ with different $\Delta z$ steps from one island to another.

## 7- Stepwise Algorithm for Numerical implementation

In this section, a step by step method is presented for the numerical evaluation of the LF in ultrawideband WDM systems. This algorithm is the logical result of the steps performed in the mathematical derivations in the previous sections. We believe this summarized stepwise algorithm could make the software numerical implementation simpler and clearer.

We assume the link that should be analyzed contains $N_s$ fiber spans and the WDM scheme Contains $N_c$ channels with center frequencies $f_{c,1}, f_{c,2}, \dots, f_{c,N_c}$. The steps on detail are as:

1- We feed eq. (1) by input launch power to each span and for each channel (frequency) and also the Raman gain profile for each fiber span. Then using set of differential equations in eq. (1), we calculate the power evolution function along each span and for each channel. This can be done using famous differential equations numerical solving methods like *Runge-Kutta* [25]. Therefore we will have $P^{(n_s)}(z,f), \forall z$ for $f \epsilon \{f_{c,1}, f_{c,2}, \dots, f_{c,N_c}\}$ and $n_s \epsilon \{1,2,\dots,N_s\}$.
2- In this step, based on equation (6), $\rho^{(n_s)}(z,f) \triangleq \frac{P^{(n_s)}(z,f)}{P^{(n_s)}(0,f)}$, we calculate $\rho^{(n_s)}(z,f)$ functions, $\forall z$, for $f \epsilon \{f_{c,1}, f_{c,2}, \dots, f_{c,N_c}\}$ and for $n_s \epsilon \{1,2,\dots,N_s\}$.
3- Having $\rho^{(n_s)}(z,f)$ from previous step, the values of $\rho^{(n_s)}\left(L_s^{(n_s)}, f\right)$ are calculated for $f \epsilon \{f_{c,1}, f_{c,2}, \dots, f_{c,N_c}\}$ and $n_s \epsilon \{1,2,\dots,N_s\}$ because we need them in the LF.
4- All possible non-null integration islands are calculated. Each calculation island is the interaction of three different WDM channels in f1-f2 plane [19]. For each integration island the following steps needs to be followed independently.
5- The value of $\vartheta^{(n_s)}(z', f_1, f_2, f_3)$, $\forall z'$ and for each span $n_s \epsilon \{1,2,\dots,N_s\}$ is calculated based on equation (19) for the current integration island.
6- The value of $\Psi^{(n_s)}(z', f_1, f_2, f_3)$, $\forall z'$ and for each span $n_s \epsilon \{1,2,\dots,N_s\}$ is calculated based on equation (22) for the current integration island.
7- $m_W$ and $N_\Psi$ are selected by user, for example we can set them to $m_W = 2$ and $N_\Psi = 10$.
8- $W(z')$ is separately calculated for each span based on equation (37) for the current integration island.
9- Equation (41) is solved exclusively for each span and $h_k^{(n_s)}(f_1, f_2, f_3)$ for $k \epsilon \{0,1,2,\dots,N_\Psi\}$ and for each span $n_s \epsilon \{1,2,\dots,N_s\}$ will be determined for the current integration island.
10- Using equation (32), $\tau_m^{(n_s)}(f_1, f_2, f_3)$ are calculated for $m \epsilon \{0,1,2,\dots,N_\Psi\}$ and for each span $n_s \epsilon \{1,2,\dots,N_s\}$ for the current integration island.
11- We can calculate the LF based on equation (34.2) for the current integration island and make the numerical integration (with respect to frequency variables) for GN/EGN calculations as we do in case of GN/EGN without Raman (equation (17)).
12- Save the calculated GN/EGN data for the current island, then go to another integration island and jump to the step 5. If no other integration island remains, go to the next step.
13- Use all the saved data of different integration island and calculate final GN/EGN result.

## 8- Conclusion

In this paper, we presented the general *Link Function* (LF) for nonlinearity assessment of fiber optic systems with GN and EGN methods. The presented LF is capable of handling both frequency dependent loss of the fiber and arbitrary power evolution for each WDM channel. We also presented an efficient approach for numerically implementing GN and EGN with the aforementioned LF function which provides the possibility of analysis of ultrawideband coherent fiber optic systems in the presence of the ISRS and distributed Raman amplification.

## 9- Acknowledgment

This work was supported by Cisco Systems through OPTSYS 2020 contract with Politecnico di Torino and by the PhotoNext Center of Politecnico di Torino. The Authors would like to thank Stefano Piciaccia and Fabrizio Forghieri from CISCO Photonics for the fruitful discussions and interactions.

## 10 - References

[1] A. Mecozzi and R.-J. Essiambre, 'Nonlinear Shannon limit in pseudolinear coherent systems,' *J. Lightwave Technol.,* vol. 30, no. 12, pp. 2011-2024, June 15th 2012.

[2] R. Dar, M. Feder, A. Mecozzi, and M. Shtaif, 'Properties of nonlinear noise in long, dispersion-uncompensated fiber links,' *Optics Express,* vol. 21, no. 22, pp. 25685--25699, Nov. 2013.

[3] P. Poggiolini, G. Bosco, A. Carena, V. Curri, Y. Jiang, F. Forghieri, 'The GN model of fiber non-linear propagation and its applications,' *J. of Lightwave Technol.,* vol. 32, no. 4, pp. 694--721, Feb. 2014.

[4] A. Carena, G. Bosco, V. Curri, Y. Jiang, P. Poggiolini and F. Forghieri, 'EGN model of non-linear fiber propagation,' *Optics Express,* vol. 22, no. 13, pp. 16335-16362, June 2014. Extended appendices with full formulas derivations can be found in the version of this paper available on www.arXiv.org.

[5] P. Serena, A. Bononi, 'A Time-Domain Extended Gaussian Noise Model,' *J. Lightwave Technol.,* vol.33, no. 7, pp. 1459-1472, Apr. 2015.

[6] A. Bononi, P. Serena, N. Rossi, E. Grellier, F. Vacondio, 'Modeling nonlinearity in coherent transmissions with dominant intrachannel-four-wave-mixing,' *Optics Express,* vol. 20, pp. 7777-7791, 26 March 2012.

[7] M. Secondini and E. Forestieri, 'Analytical fiber-optic channel model in the presence of cross-phase modulations,' *IEEE Photon. Technol. Lett.,* vol. 24, no. 22, pp. 2016-2019, Nov. 15$^{th}$ 2012.

[8] P. Johannisson, M. Karlsson, 'Perturbation analysis of nonlinear propagation in a strongly dispersive optical communication system,' *J. Lightwave Technol.,* vol. 31, no. 8, pp. 1273-1282, Apr. 15, 2013.

[9] R. Dar, M. Feder, A. Mecozzi, M. Shtaif, 'Pulse collision picture of inter-channel nonlinear interference noise in fiber-optic communications,' *J. Lightwave Technol.,* vol. 34, no. 2, pp. 593-607, Jan. 2016.

[10] P. Poggiolini, Y. Jiang, 'Recent Advances in the Modeling of the Impact of Nonlinear Fiber Propagation Effects on Uncompensated Coherent Transmission Systems,' tutorial review, *J. of Lightwave Technol.,* vol. 35, no. 3, pp. 458-480, Feb. 2017.

[11] Namiki, Shu, et al. "Challenges of Raman amplification." *Proceedings of the IEEE* 94.5 (2006): 1024-1035.

[12] Poggiolini, Pierluigi. "A generalized GN-model closed-form formula." *arXiv preprint arXiv:1810.06545* (2018).

[13] D. Semrau, R. I. Killey, P. Bayvel, 'A Closed-Form Approximation of the Gaussian Noise Model in the Presence of Inter-Channel Stimulated Raman Scattering,' *www.arXiv.org,* paper arXiv:1808.07940, Aug. 23rd 2018.

[14] Zefreh, Mahdi Ranjbar, and Pierluigi Poggiolini. "A Real-Time Closed-Form Model for Nonlinearity Modeling in Ultra-Wide-Band Optical Fiber Links Accounting for Inter-channel Stimulated Raman Scattering and Co-Propagating Raman Amplification." *arXiv preprint arXiv:2006.03088* (2020).

[15] RosaBrusin, Ann Margareth, Uiara Celine De Moura, Vittorio Curri, Darko Zibar, and Andrea Carena. "Introducing Load Aware Neural Networks for Accurate Predictions of Raman Amplifiers." *Journal of Lightwave Technology* (2020).

[16] Mochizuki, Kiyofumi. "Optical fiber transmission systems using stimulated Raman scattering: Theory." *Journal of lightwave technology* 3.3 (1985): 688-694.

[17] Tariq, Salim, and Joseph C. Palais, 'A computer model of non-dispersion-limited stimulated Raman scattering in optical fiber multiple-channel communications,' *J. of Lightwave Technology,* vol. 11, no. 12, pp. 1914-1924, 1993.

[18] Alberto Bononi, Ronen Dar, Marco Secondini, Paolo Serena and Pierluigi Poggiolini, 'Fiber Nonlinearity and Optical System Performance,' Chapter 9 of the book: *Springer handbook of Optical Networks,* editors Biswanath Mukherjee, Ioannis Tomkos, Massimo Tornatore, Peter Winzer, Yongli Zhao, ed. Springer, 2020. ISBN 978-3-030-16250-4.

[19] Zhou, Xiang, and Chongjin Xie. *Enabling technologies for high spectral-efficiency coherent optical communication networks*. John Wiley & Sons, 2016.

[20] Poggiolini, Pierluigi. "The GN model of non-linear propagation in uncompensated coherent optical systems." *Journal of Lightwave Technology* 30.24 (2012): 3857-3879.